
\input harvmac

\def\t {\tau}

\def \p {\phi}
\def \ha {\half}
\def \ov {\over}

\def \four{{\textstyle {1\ov 4}}}
\def \a {\alpha}
\def \lr { \lref}
\def\ep{\epsilon}

\def \r {\rho}
 \def\m{\mu}\def\n {\nu}\def\l
{\lambda}

\def \P {\Phi}

\def   \td {\tilde }
\def \k {\kappa}
\def \lr { \lref}

\gdef \jnl#1, #2, #3, 1#4#5#6{ { #1~}{ #2} (1#4#5#6) #3}

\def\np {  Nucl. Phys. }
\def \pl { Phys. Lett. }

\def \prl { Phys. Rev. Lett. }
\def \pr  { Phys. Rev. }

\def \ijmp { Int. J. Mod. Phys. }

\baselineskip8pt
\Title{\vbox
{\baselineskip 6pt{\hbox{  }}{\hbox
{Imperial/TP/95-96/6 }}{\hbox{hep-th/9510173}} {\hbox{
   }}} }
{\vbox{\centerline {On  $SO(32)$ heterotic -- type I superstring }
 \centerline {duality in ten dimensions}
 }}

\vskip -20 true pt






\medskip
\centerline{   A.A. Tseytlin\footnote{$^{\star}$}{\baselineskip8pt
e-mail address: tseytlin@ic.ac.uk}\footnote{$^{\dagger}$}{\baselineskip8pt
On leave  from Lebedev  Physics
Institute, Moscow.} }

\smallskip\smallskip
\centerline {\it  Theoretical Physics Group, Blackett Laboratory,}
\smallskip

\centerline {\it  Imperial College,  London SW7 2BZ, U.K. }
\bigskip\bigskip
\centerline {\bf Abstract}
\medskip
\baselineskip10pt
\noindent
\medskip
We provide some additional evidence in favour of the
strong -- weak coupling duality between the $SO(32)$ heterotic
and type I superstring theories by comparing terms quartic in the
gauge field strength in their low-energy effective actions.
We argue that these terms should not receive
higher-loop string corrections so that
duality should relate the leading-order perturbative coefficients
in the two theories. In particular, we demonstrate that
the coefficient of the $F^4$-term in the one-loop (torus) part
of the $SO(32)$ heterotic string action is exactly the same
as the coefficient of the $F^4$-term in the tree-level (disc) part
of the type I action.

\Date {October 1995}

\noblackbox
\baselineskip 14pt plus 2pt minus 2pt
\def \tr {{ \rm tr  }}
\def \k {\kappa}

\lr \met {R.R. Metsaev and A.A. Tseytlin, \np B298 (1988) 109.}

\lr \chap {A.H. Chamseddine, \np B185 (1981) 403;
E. Bergshoeff, M. de Roo, B. de Wit and P. van Nieuwenhuizen, \np B195 (1982)
97;
G.F. Chaplin and N.S. Manton, \pl B120 (1983) 105.}
\lr \witten {E. Witten, \np B443 (1995) 85. }
\lr \dabgib {A. Dabholkar, G.W. Gibbons, J. Harvey and F. Ruiz Ruiz,  \np
B340 (1990) 33;
A. Dabholkar and  J. Harvey,  \prl
63 (1989) 478.
}

\lr\dab{A. Dabholkar, \pl B357 (1995) 307,
hep-th/9506160.}
\lr \hul{
C.M. Hull, \pl B357 (1995) 545, hep-th/9506194.}

\lr \ft { E.S. Fradkin  and A.A. Tseytlin, \pl B160 (1985) 69.  }
\lr \ftt { E.S. Fradkin  and A.A. Tseytlin, \np B227 (1983) 252.  }
\lr \ler { W. Lerche, \np B308 (1988) 102.}
\lr\lerr{W. Lerche, B.E.W. Nilsson and  A.N. Schellekens, \np B289 (1987)
609;
W. Lerche, B.E.W. Nilsson, A.N. Schellekens and N.P. Warner, \np B299 (1988)
91.}

\lr\gross{ Y. Cai  and C. Nunez, \np B287 (1987) 279; Y. Kikucki  and C.
Marzban, \pr D35 (1987) 1400.}
\lr\grosss{ D.J.  Gross and J.H. Sloan, \np B291 (1987) 41. }
\lr\elli { J. Ellis, P. Jetzer and L. Mizrachi, \np B303 (1988) 1. }
\lr\ell { J. Ellis and L. Mizrachi, \np B302 (1988) 65. }

\lr \abe { M. Abe, H. Kubota and N. Sakai, \np B306 (1988) 405. }
\lr \gs{ M.B. Green  and J.H. Schwarz, \pl B149 (1984) 117;
\pl B151 (1985) 21; \np B255 (1985) 93. }
\lr \gro{D.J. Gross  and E. Witten, \np B277 (1986) 1.}
\lr \tse { A.A. Tseytlin, \np B276 (1986) 391. }
\lr \callan {C.G. Callan, C. Lovelace, C.R. Nappi and S.A. Yost, \pl B206
(1988) 41;
\np B308 (1988) 221.}
\lr \elw {U. Ellwanger, J. Fuchs and M.G. Schmidt, \np B314 (1989) 175.}
\lr \green {M.B. Green, J.H.  Schwarz and E.  Witten, {\it Superstring Theory}
(Cambridge U.P., 1988).}
\lr \abo{ A. Abouelsaood, C. Callan, C. Nappi and S. Yost, \np { B280 }
 (1987) 599.}
\lr \aob{ E. Bergshoeff, E. Sezgin, C.N. Pope and P.K. Townsend, \pl B188
(1987) 70. }
\lr \frt{ E.S. Fradkin and A.A. Tseytlin, \pl {
B163  }
(1985) 123.}

\lr \abb {N. Sakai and M. Abe, Progr. Theor. Phys. 80 (1988) 162.}
\lr \grom{D.J. Gross, J.A. Harvey, E. Martinec and R. Rohm, \np B256 (1985)
253; B267 (1986) 75.}
\lr \mrt {R.R. Metsaev, M.A. Rahmanov and A.A. Tseytlin, \pl B193 (1987) 207;
A.A. Tseytlin, \pl B202 (1988) 81.}

\lr \bos{J.A. Shapiro and C.B. Thorn, \pr D36 (1987) 432;
J. Dai and J. Polchinski, \pl B220 (1989) 387.}

\lr \cai {  J. Polchinski and Y. Cai, \np B296 (1988) 91.}

\lr \green {M.B. Green, J.H.  Schwarz and E.  Witten, {\it Superstring Theory}
(Cambridge U.P., 1988).}

\lr\gibbb{G.W. Gibbons and D.A. Rasheed, ``Electric-magnetic duality rotations
in non-linear electrodynamics", DAMTP-R-95-46,
  hep-th/9506035;
``SL(2,R) invariance of non-linear electrodynamics coupled to an axion and a
dilaton", DAMTP-R-95-48,
hep-th/9509141.  }
\lr \pow{J. Polchinkski and E. Witten, ``Evidence for heterotic -- type I
string duality", IASSNS-HEP-95-81, NSF-ITP-95-135, hep-th/9510169.}

\lr\roo {M. de Roo, H. Suelmann and A. Wiedemann, \np B405 (1993) 326; H.
Suelmann, ``Supersymmetry and string effective actions", Ph.D. thesis,
Groningen, 1994.   }

\lr \sak {N. Sakai and Y. Tanii, \np B287 (1987) 457.}
\lr \mei{ M. Abe, H. Kubota and N. Sakai, \pl B200 (1988) 461;
K.A. Meissner, J. Pawelczyk and S. Pokorski, \pr D38 (1988) 1144.}
\lr \yas {O. Yasuda, \pl B218 (1989) 455. }
\lr \miz { L. Mizrachi, \np B338 (1990) 209.}
\lr \yass {O. Yasuda, \pl B215 (1988) 306.  }
\lr \ieng{A. Morozov, \pl B209 (1988) 473;  R. Iengo  and C.-J. Zhu, \pl B212
(1988) 313.}
\lr \nils{ B.E.W. Nilsson and A.K. Tollsten, \pl 169 (1986) 369; R. Kallosh,
Phys. Scr. T15 (1987) 118.  }
\lr \tsss{  A.A. Tseytlin, \ijmp A5 (1990) 589.}
\lr \alw{L. \'Alvarez-Gaum\'e and E. Witten, \np B234 (1983) 269.     }

The $D=10$ supersymmetry implies that the type I superstring
and the  $SO(32)$ heterotic string  theories  \refs{\gs,\green,\grom}
should be described by the same leading-order
low-energy (plus anomaly-cancelling terms) action --  the $N=1, D=10$
supergravity plus supersymmetric Yang-Mills action \chap.
Thus the leading-order terms in the corresponding
 actions  should be related by a field
redefinition.
As was observed in \witten,
this  field transformation involves reversing  the sign of the dilaton
and thus interchanges  the weak and
strong  coupling regimes of the two theories.
This suggests \witten\ that the $SO(32)$ heterotic and type I string theories
in ten dimensions
are  dual to each other in the sense that a strong-coupling
region  of one theory can be  described
by dynamics of solitonic states which is equivalent
to the weak-coupling  dynamics of elementary states of the other.
Such   duality relation  is
supported by the fact that the fundamental string solution
of the heterotic string \dabgib\  appears as a
 `soliton' of   type I theory
when rewritten in terms of the  type I string
metric and dilaton \refs{\dab,\hul}.

To check this duality  further one may try to
compare higher-order  terms in the two effective actions.
They should   be equivalent at each order  in  derivative  expansion
but exactly in string coupling.
As we shall show below, it is  indeed possible  to
establish a correspondence between  the  gauge field
$F^4$-terms
in the
heterotic and type I effective actions which is in agreement with  the
duality of the two theories.


The leading-order  terms  in the two effective actions have the form
\eqn\act{ S_{het}=
 \int d^{10} x \sqrt G\  \big\{ c_0  e^{-2\p} \big[R  + 4 (\del \p)^2 - {1\ov
12} \hat H^2_{\m\n\l}
+ \  {1\ov 8}  \tr F^2_{\m\n} \big]
 + ... \big\}\  , \   }
\eqn\acto{ S_{typeI}=
  \int d^{10} x \sqrt {G'}\{ c_0  e^{-2\p'}   [R'  + 4 (\del \p')^2] - {1\ov
12} c'_0  \hat H^2_{\m\n\l} } $$
+ \  {1\ov 8} a_0  e^{-\p'} \tr F^2_{\m\n}
 +  ... \} \ . $$
We shall use primes to indicate the fields of the type I theory.
The trace $\tr$ is  in the fundamental representation of
$SO(32)$,  $\tr(T_a T_b)=-2\delta_{ab}$
($A_\m= A^a_\m T_a$,  $ \ F_{\m\n} =  \del_\m A_\n -\del_\n A_\m +
[A_\m, A_\n])$ and  $\hat H$ contains the  Chern-Simons terms,
$\hat H = dB  + \ha  \omega_3(A) + ...$, \ $\omega_3(A)= \tr (AdA + {2\ov 3}
A^3)   $.
The string coupling is, as usual, absorbed into
the constant part of  the dilaton.
Similarly, we assume
(as it is natural from the point of view of the world-sheet action)
that
 the string tension  is absorbed into the metric and  the 2-form field
which thus have the
 dimension $cm^{-2}$ (all  the parameters in \act,\acto\ are then
dimensionless
while  the tensors have the
geometrical dimension, $[T_{\m_1...\m_n}]= cm^{-n}$).
In a standard  perturbative phase,
\eqn\flat{ G_{\m\n} = {1\ov \a'} \delta_{\m\n} + ... \ ,
\ \ \ \  G'_{\m\n} = {1\ov \a'_I} \delta_{\m\n} + ... \ ,  }
where $\a'$ and $\a'_I$ are the  inverse tensions of the heterotic and type I
 theories.
The terms in \act\ are  the tree-level (sphere) ones
while the three groups of terms in \acto\
with different  powers of $e^{\p'}$
correspond to the sphere, annulus\foot{Since the antisymmetric tensor of the
type I theory appears in the RR-sector,
heuristically one may think of   its kinetic term as
  originating
 from the annulus diagram (each of the two non-local spin operators cuts
 a disc out of the sphere).
  We follow  \witten\ (cf.\callan)
and use $B_{\m\n}$ redefined by the  factor of $e^{\p'}$
so that $\hat H$ does not contain the
dilaton dependence (both in  the heterotic  and type I  cases). } and disc
diagrams.

The  actions  \act\ and  \acto\
are related by \witten\foot{The consistency (supersymmetry)
requires  that in \acto\  $c_0'= a_0 = c_0$
(though this was not checked directly, cf.\callan).
The coefficient   $a_0$ is proportional to the value of the dilaton  tadpole
on the disc  and determines the ratio of the Yang-Mills and gravitational
couplings in  the type I theory \refs{\cai, \abb}\ (see below).
Note  also that  one may use alternative normalisations
introducing    an extra constant
shift in the relation between the two dilatons, $\p'=-\p + p,$  while
keeping  the two tensions in \flat\ the same.}
\eqn\red{ G'_{\m\n} = e^{-\p} G_{\m\n}\ , \ \ \
\p'=-\p  \ , \ \ \ \   B'_{\m\n}=B_{\m\n}\ , \ \ A'_\m = A_\m \ .  }
The  higher-order  $F^4$-terms  in the  two  actions have the following
structure:
\eqn\actt{ \Delta S_{het}  =
  \int d^{10} x \sqrt G \big[   f(\p)\tr F^4 +   h (\p) \tr F^2 \tr F^2
 \big]\  ,   }
\eqn\atto{ \Delta S_{typeI} =
  \int d^{10} x \sqrt{ G' }\big[   f'(\p')\tr F^4 +   h' (\p') \tr F^2 \tr F^2
 \big] \ ,   }
where the indices of the four $F_{\m\n}$-factors  are contracted
with  $\tau^{\m_1...\m_8}$
(which is the 10-dimensional extension of the 8-dimensional
light-cone gauge `zero-mode' tensor \green\
with $\epsilon^{\m_1...\m_8}$-term omitted)
 built out of $G^{\m\n}$  in \actt\
and $G'^{\m\n}$ in \atto, e.g.,
\eqn\ffff{ \tr F^4\equiv
 \tau^{\m_1\n_1 ... \m_4\n_4}\tr (F_{\m_1\n_1} F_{\m_2\n_2} F_{\m_3\n_3}
F_{\m_4\n_4})} $$  = 16 \tr\big(F^{\m\n}F_{\r\n}F_{\m\l}F^{\r\l} +
\ha  F^{\m\n}F_{\r\n}F^{\r\l} F_{\m\l} $$ $$
- \  \four F^{\m\n}F_{\m\n}F^{\r\l}F_{\r\l}
-{\textstyle {1\ov 8}} F^{\m\n}F^{\r\l}F_{\m\n}F_{\r\l} \big). $$
The functions $f,h$ and $f',h'$ are, in general, expected to have the
following perturbative expansions
(for small $e^\p$ and small $e^{\p'}$ respectively)
\eqn\hh{ f (\p) = b_0 e^{-2\p} + b_1 + b_2 e^{2\p}  + ... + b_n e^{2(n-1)\p} +
... \ , }
\eqn\hhh{ h (\p) = c_0 e^{-2\p} + c_1 + c_2 e^{2\p}  + ... + c_n e^{2(n-1)\p} +
... \ , }
and
\eqn\heh{ f'(\p')  = b'_0 e^{-\p'} + b'_1 + b'_2 e^{\p'}  + ... + b'_n
e^{(n-1)\p'} + ... \ , }
\eqn\hehh{ h'(\p') = c'_0 e^{-\p'} + c'_1 + c'_2 e^{\p'}  + ... + c'_n
e^{(n-1)\p'} + ... \ . }
Some  of the leading-order coefficients were computed in the past.
In the heterotic string theory
the tree-level (sphere) coefficients are \refs{\gross,\grosss,\elw}\
$ b_0=0, \ c_0\not=0$  and the  one-loop (torus)
coefficients in the $SO(32)$ heterotic theory
are \refs{\elli,\abe,\ler} $b_1\not=0, \ c_1=0$.
In the type I theory the   tree-level  (disc) coefficients are
 \refs{\tse,\gro} $b'_0\not=0, \ c'_0=0$.
While the fact that $c'_0=0$ is trivial (disc has just one boundary),
 the reason why $ b_0=0$ and $c_1=0$ (in the $SO(32)$ heterotic theory)
was not understood before.
As we shall see, these values are
indeed `explained'
 by the duality.\foot{The
 double trace  ($\tr F^2 \tr F^2$) term does appear (i.e. $c_1\not=0$)
 in the  one-loop action of the $E_8\times E_8$ heterotic string theory \abe\
  which  indeed cannot  be directly
related to the type I theory.}

Applying the duality transformation  \red\
to  the type I action \atto, i.e. expressing it in terms of the heterotic
string metric and dilaton,  we get
\eqn\attos{ \Delta S_{typeI}  =
  \int d^{10} x \sqrt G \big[   \td  f(\p)\tr F^4 +  \td  h (\p) \tr F^2 \tr
F^2
 \big] \ ,
}
where
\eqn\cee{
\td  f(\p) \equiv e^{-\p} f'(-\p) = b'_0  + b'_1  e^{-\p}
+ b'_2 e^{-2\p}  + ...  + b'_n e^{- n \p} + ... \ , }
\eqn\ceee{
\td  h(\p) \equiv e^{-\p} h'(-\p) = c'_0  + c'_1  e^{-\p}
+ c'_2 e^{-2\p}  + ...  + c'_n e^{- n \p} + ... \ .  }
The  duality  equivalence  should imply that\foot{Note that in addition
to changing  the sign of the dilaton the transformation between
the coupling functions of the two theories involves
rescaling by the factor $e^{-\p}$ in \cee,\ceee. This
is  a reflection of the fact that not only the dilaton (or string coupling)
but also the metric (or string tension)
is transformed.}
\eqn\duud{ \td  f(\p) = f(\p)  \ , \ \ \ \ \
\td  h(\p) = h(\p)  \ .  }
Since $f(\p), h(\p)$  are known only for large negative
$\p$ (small heterotic string coupling) while
$f'(-\p), h'(-\p)$  (and thus  $\td f(\p), h (\p)$)
-- for large positive $\p$ (small type I string coupling)
one cannot, in general,  test duality  just  by comparing the
coefficients of  the
 few terms in the above  formal expansions which  happen to have the
same dilaton dependence.\foot{I am grateful to E. Witten for emphasizing this
point  to me.  Note also  that it is the absence of perturbative higher-loop
corrections
to the leading-order terms in \act,\acto\ that makes
legitimate  the consideration of the duality transformation \red\ between them
(in addition, one  assumes  that the massless part of strong-coupling solitonic
spectrum of one theory is the same as the perturbative weak coupling spectrum
of the other, as is indeed implied by supersymmetry).}
It would be possible to  check   \duud\
using   perturbation-theory results  only if most of the coefficients
in \hh,\hhh,\heh,\hehh\ were zeroes. Indeed, \duud\ would be satisfied
provided
\eqn\sat{ b'_0= b_1\ , \ \ \ b_1'=0\ , \ \ \
 b'_2=b_0\ , \ \ \
b_3'=b_4'=...=0\ , \ \  \ b_2=b_3=...=0\ , }
\eqn\sats{c'_0= c_1\ , \ \ \ c_1'=0\ , \ \ \
 c'_2=c_0 \ , \ \ \
c_3'=c_4'=...=0\ , \ \  \ c_2=c_3=...=0\ . }
Note that  the absence of
double-trace term in tree-level part of type I action  ($c_0'=0$) is thus
equivalent
by duality to the known
absence of such term in the 1-loop part of the   $SO(32)$ heterotic string
action ($c_1=0$).
Also,  the  absence of the single-trace
term in the tree-level heterotic  action ($b_0=0$)
implies that there should be no such term coming from
the genus $-1$ diagrams in type I theory.
At the same time,
the presence of the double-trace  term
in the tree-level heterotic action ($c_0$) is
not surprising given that the  duality relates  it to  the
 genus $-1$ correction ($c'_2$)
in type I theory  which
originates  from  diagrams with  $m\leq 3$
boundaries.

Taking into account the known perturbative results
($b_0=c_1=c_0'=0$),
the conditions  \sat,\sats\  thus require that
\eqn\req{ f(\p) = b_1 \ , \ \ \  f'(\p') = b_0' e^{-\p'}\ , \ \ \
  b_1=b'_0\ , }
\eqn\reqq{ h(\p) = c_0 e^{-2\p}  \ , \ \ \  h'(\p') = c'_2 e^{\p'}\ , \ \ \
  c'_2=c_0 \ .  }
In what follows we shall consider only  the $\tr F^4$ coupling functions  $f$
and $f'$ and try to justify \req,\sat\ and thus the duality relation \duud.
We shall first argue
 that $f$ and $f'$  should  not contain higher-loop
corrections  and then demonstrate the equality of the coefficients
$b_1$ and $b_0'$.

The reason why the 1-loop $\tr F^4$-term
in heterotic string theory  should  not receive higher-loop corrections
is its  direct  connection \ler\  with
  the anomaly-cancelling term
$\epsilon^{\m_1 ...\m_{10}} B_{\m_1\m_2} \tr
(F_{\m_3\m_4}F_{\m_5\m_6}F_{\m_7\m_8} F_{\m_9\m_{10}}) $
\refs{\gs,\lerr,\elli,\ell}.
The  derivation  \ler\
of this term based on light-cone Green-Schwarz superstring
 partition function representation  for
the effective action
on the torus \ft\
 is closely related  to
the chiral anomaly index  computation \lerr\
and can probably  be generalised to
higher loops,  giving  zero result
in view of
 the expected absence of higher-loop contributions to the anomaly cancellation
condition.\foot{The relation between  the 4-vector parity-even
and  the  four vector -- one antisymmetric tensor
 parity-odd terms
was noticed at the level of the 1-loop
string amplitudes \refs{\elli,\ell}. The close connection  between the
calculation
of the anomaly index and the one-loop $O(R^4,R^2F^2,F^4)$ term
 in  the heterotic string effective action
was suggested as an indication that this term does not receive higher
 string loop corrections \ler.
This connection is also apparent from the analysis of \yas.
It should be emphasised  that the condition of
preservation of supersymmetry is crucial for this non-renormalisation.
For example, the two-loop
4-point amplitude as given in \ieng\
may look non-vanishing
when expanded to $k^4$-order in momenta;
the supersymmetry requires, however,  that the
coefficient multiplying the kinematic factor $K\sim k^4 $
must vanish under a proper computational procedure  consistent with
the supersymmetry and the anomaly cancellation (see also \yass).}

It is  the $D=10$ supersymmetry combined with the absence
 of higher loop
contributions to the anomaly-cancelling terms
that  rules out higher-loop  corrections to $\tr F^4$ (and also to $\tr F^2 \tr
F^2$).
Indeed, the two  super-invariants
including the $F^4$-structures  contain  the following bosonic terms \roo\
\eqn\super{
I_1= \tr F^4 - \four \epsilon^{\m_1 ...\m_{10}} B_{\m_1\m_2} \tr
(F_{\m_3\m_4}F_{\m_5\m_6}F_{\m_7\m_8} F_{\m_9\m_{10}}) \ , }
$$
I_2= \tr F^2 \tr F^2  - \four \epsilon^{\m_1 ...\m_{10}} B_{\m_1\m_2} \tr
(F_{\m_3\m_4}F_{\m_5\m_6})\tr (F_{\m_7\m_8} F_{\m_9\m_{10}}) \ .
$$
Thus the absence of the higher-loop contributions
 to the coefficients of the  $BF^4$-terms \refs{\yas,\miz}
implies that the  coefficients of
$I_1$ and $I_2$ are not renormalised.\foot{A heuristic way of understanding the
connection beween $F^4$ and $BF^4$ terms
 is the following.
The  light-cone gauge  `zero-mode' tensor  which is present
 in the 4-point amplitude \green\
$t^{\m_1...\m_8}_8 = \t^{\m_1...\m_8} - \ha \ep^{\m_1...\m_8} $
 may be given the following   10-dimensional
generalisation:
$t^{\m_1...\m_8}_{10} = \t^{\m_1...\m_8} - \four B_{\l\r}
\ep^{\l\r\m_1...\m_8}$
(assuming that in  the light-cone gauge $B_{uv}=1, \ F_{uv}=0$).
Then the combinations  that appear in the super-invariants \super\ are
just $\tr F^4$ and $\tr F^2 \tr F^2$ with $\t_8$ replaced by
$t_{10}$.
  This is also consistent with the structure of the corresponding string
amplitudes
\refs{\elli,\ell} (i.e. correlators of the
vertex operators)
 which of course should satisfy the requirement of
the linearised supersymmetry. }

The absence of loop  corrections to
the tree-level $\tr F^4$-term in the $SO(32)$  type I theory
is also implied by   the supersymmetry and  the
anomaly cancellation
(here
the anomaly-cancelling term $B\tr F^4$  comes only from the tree-level disc
diagram \gs).
More explicitly, the  diagrams which may contribute to this term
have the topology of a disc with four external
 legs at the boundary and insertions of holes, crosscaps and handles
at the interior.  Let us start with  the 1-loop
correction given by the sum of the annulus and the M\"obius strip.
The corresponding  amplitude  has the following structure
\gs\
\eqn\ampl{ A_4 = 16 K \int^1_{-1} {d\l \ov \l} \int [d\nu] \prod_{i<j}
[B(\n_i-\n_j, \l)]^{2\a'_I k_i\cdot k_j} \ , }
where $K= - \four st \zeta_1\cdot \zeta_2 \zeta_3\cdot \zeta_4 + ... $
is the standard kinematic factor
and the integral over $\l$ is defined using the principal value prescription.
$K$ itself gives already the right  momentum structure to reproduce the $\tr
F^4$-term
so  that in
order to compute its   1-loop coefficient $b_1'$
one should  set $k_i=0$ in the integrand of \ampl.
The latter then vanishes because of  the principal value prescription \gs\
(cancellation  between the annulus and the M\"obius strip contributions), i.e.
  for the same reason
why there is no non-trivial dilaton tadpole (or vacuum partition function)
in the  $SO(32)$
type I theory (see also \cai).
The fact that $b_1'=0$
is important for the duality to work since
there is obviously no place for a perturbative
$e^{-\p}$-correction in the heterotic string theory, cf.\cee,\duud.\foot{It
is interesting to note that the vanishing of the coefficient
of the 1-loop $\tr F^4$ term is a string-theory effect:
there {\it is}  (quadratically divergent) 1-loop $\tr F^4$ term
in the $D=10$ super Yang-Mills theory \ftt. It  can be reproduced
from the string theory
by taking  the $D=10$  field-theory  limit ($\a'_I \to 0$ for  fixed UV cutoff,
i.e.
$\l \to 1$)
in the type I amplitude \ampl\ \met.}

The same reasoning is readily extended to higher-loop type I diagrams:
the  right kinematic structure  needed
to reproduce $F^4$ is always  given by
 $K$ (which comes, e.g., from the integral over the  zero modes of
GS fermions in the partition function representation of the effective action
 \refs{\tse,\gro}) while the remaining part of the diagram
should be taken  at zero momentum and thus
must vanish  being just  the sum of vacuum diagrams.
We thus learn that $b'_2=b'_3=...=0$.
Note, in particular, that  the fact that $b'_2=0$ combined with duality \sat\
provides an `explanation' why $b_0=0$, i.e. why there is no  $\tr F^4$-term in
the tree-level heterotic string action \refs{\gross,\grosss}.

The core of the argument
that  $b'_1=b'_2=...=0$ is the  condition
of  finiteness  (related to anomaly cancellation)
of  the $SO(32)$ type I theory. This  is in  correspondence with
 the above  claim
about  similar   connection  between  the automatic higher-loop
anomaly cancellation and
the vanishing of higher-loop corrections to $\tr F^4$ ($b_2=b_3=...=0$)
in the heterotic case.

The further   check
of the  equivalence  between  the two  $SO(32)$ theories
is provided by the
 precise   equality  of
 the  type I  tree-level  coefficient $b_0'$
and the heterotic one-loop coefficient $b_1$ in \req.
To demonstrate that $b_0'=b_1$
 it is useful
to represent   the actions \act,\actt\ and \acto,\atto\
in terms of the  usual
Einstein-frame metric $g_{\m\n}$
 and the non-constant part $\P$
of the dilaton field ($\a' G_{\m\n} = e^{\p/2} g_{\m\n} ,
\ \  \p=\p_0 + \P$)  and the  10-dimensional
gravitational and Yang-Mills coupling constants.
We get for the heterotic string action
\eqn\acte{ S_{het}=
 \int d^{10} x \sqrt g \  \{ -{1\ov 2\k^2}
\big[R (g)  -{1\ov 2}  (\del \P)^2 - {1\ov 12} e^{-\P}  \hat H^2_{\m\n\l}\big]}
$$
-\     {1\ov 8 g^2_{10}}  e^{-\P/2}  \tr F^2_{\m\n}  + ...
 +  b_1  e^{\P/2} \tr F^4  + ...  \}\  ,   $$
where the gravitational $\k$ and  the Yang-Mills $g_{10}$ couplings are related
to the  dimensionless heterotic string coupling  $g $
by \grom\foot{Note  also that \red\ implies that
the two string tensions  are related by  $\ \a'= e^{- \p_0}\a'_I, \ \
\a' \a'_I = \ha \k$ (see also  \dab).}
 \eqn\cve{ \k^2= 4\a'^4 g^2\ , \ \ \ \ \ g^2_{10} = 2\a'^3 g^2\ , \ \ \ \ g =
e^{\p_0}\ ,  }
 and the one-loop (torus) coefficient  $b_1$ is  given by
\refs{\elli,\abe}
\eqn\coe{ b_1 =  - {\a'^3 \ov 3 \times 2^{11} \pi^5 } \big({g\ov \k}\big)^2
=  - {1 \ov 3 \times 2^{12} \pi^5 } \big({g_{10}\ov \k}\big)^2 \ . }
The type I action takes the form ($\p'=\p'_0 + \P', \ \P'=-\P$,  see \red)
\eqn\actos{ S_{typeI}=
 \int d^{10} x \sqrt g \  \{ -{1\ov 2\k'^2}
\big[R (g)  -{1\ov 2}
 (\del \P')^2 - {1\ov 12} e^{\P'}  \hat H^2_{\m\n\l}\big]  } $$
-  \  {1\ov 8 g'^2_{10}}  e^{\P'/2}  \tr F^2_{\m\n}  +
   b_0' e^{-\P'/2} \tr F^4  + ...  \} , \   $$
where  the couplings $\k'$ and $g'_{10}$ are related to the closed and open
superstring string couplings  $g'$ and $g_o$
by \abb\foot{See also \cai;  a similar relation
in  the bosonic string case was derived in  \bos. In our notation the
normalisation coefficient
of the $SO(32)$ gauge group generators in the fundamental representation is
$T(F)= \ha$.}
 \eqn\cvee{ \k'^2= 4\a'^4_I g'^2 , \  \ g'^2_{10} = 2\a'^3_I g^2_o ,
\    \   g^4_o = (2\pi)^7 g'^2  , \ \
   g'^4_{10} = (2\pi)^7\a'^2_I \k'^2  , \   \ \
g'= e^{ \p_0'} . }
The tree-level (disc) open string coefficient $b_0'$ is \refs{\tse,\gro}
\eqn\coff{  b'_0 =  - {(2\pi \a'_I)^2 \ov 3 \times 2^7 g'^2_{10}}=
 - {1 \ov 3 \times 2^{12} \pi^5 } \big({g'_{10}\ov \k'}\big)^2 \ . }
It is  exactly the same as $b_1$ \coe\ provided we set
 $ \k=\k' ,  \ g_{10} = g'_{10}.$

The coincidence of the two coefficients $b_1$ and $b'_0$
may seem  quite remarkable given the very different nature
of the string diagrams  they are extracted from.
In view of the above discussion, it
 may, however, be considered
 as being just a consequence
of the $D=10$  supersymmetry and the consistency (anomaly cancellation)
of the two string  theories.
Indeed,  the two low-energy theories are both equivalent
to the $D=10$ supergravity + Yang-Mills  theory.
The latter $SO(32)$ field theory
is free \gs\ from  1-loop anomalies \alw\   provided
one adds the anomaly-cancelling terms $BF^4$
with specific coefficients (determined just by the field theory, i.e.
by the structure of the  massless
particle 1-loop diagrams).
Then the supersymmetry
 demanding that   $BF^4$ and $F^4$ should  appear as parts of the same
super-invariant \super\
implies also  that the coefficient of
the $F^4$-term is    fixed uniquely by the
low-energy field theory and thus {\it must}  be the same
in the two string theories.

The above discussion   suggests that the  local terms in the
two effective actions \acte\  and \actos\
which  appear only at  specific  orders of perturbation theory
or   related  to such terms   by the supersymmetry
are  indeed the same up to $\P'=-\P$.\foot{It should be noted that  in addition
to the local terms,
the  massless superstring effective actions contain also
non-local terms which are non-analytic in momenta.
We define the  string effective action
as the one the tree-level amplitudes of which reproduce the full loop-corrected
string amplitudes for massless states. The non-analyticity of the low-energy
expansion is due to loops of  massless
string states  which
must necessarily be included in order to have a well-defined  (finite,
anomaly-free)
effective action (see \refs{\tsss}).}
One may  try to  use this equivalence
to  learn  more about the structure of certain
terms in one action by starting with their analogs  in
the other one (which may be easier to
compute directly).
For example, let us
consider the special  case when the
 vector field $A_\m$  is  taken to be in
an  abelian $SO(2)$  subgroup of $SO(32)$.
Then the $\del F$-independent  part of the  tree-level (disc)
$e^{-\p'}$-term  in  the open superstring action \acto\
 is given by the Born-Infeld action \refs{\frt,\mrt,\abo,
\aob,\callan}\foot{Note that
$\sqrt{{\rm det} (\delta^\m_\n  +
2\pi  G'^{\m\l} F_{\l\n})} - 1 = \four (2\pi \a'_I)^2 \big\{  F^2_{\m\n}  -
 \ha (2\pi \a'_I)^2 [F^4 - \four (F^2)^2]\big\}  + O(F^6).$
The $F^2,F^4$-terms are in agreement with \acto,\actos,\coff\ ($a_1=
2a_0/\pi^2$).
Note also that in contrast to the oriented
bosonic  string case (see e.g. \met),
 in the non-oriented type I case the disc term in the action
does not depend on  $B_{\m\n}$.}
 $$ S_{typeI}=
 \int d^{10} x \sqrt {G'} \{ c_0 e^{-2\p'}   [R'  + 4 (\del \p')^2] - {1\ov 12}
c_0'  \hat H^2_{\m\n\l}  $$
\eqn\aco{  +\   a_1  e^{-\p'}  [ \sqrt{{\rm det}  (\delta^\m_\n  + 2\pi
G'^{\m\l} F_{\l\n})} -
1 ]  + ...\} \ . }
Rewritten in terms of the heterotic string metric and dilaton  in  \red\
   this
action  becomes (cf.\act,\actt)
$$S_{het} (G,\p)=S_{typeI}(G',\p')=
  \int d^{10} x \sqrt {G} \{ c_0 e^{-2\p}   [R +  4 (\del \p)^2 - {1\ov 12}
\hat H^2_{\m\n\l}] $$
 \eqn\acov{
+\  a_1   e^{-4\p}  [ \sqrt{{\rm det} (\delta^\m_\n  + 2\pi  e^\p G^{\m\l}
F_{\l\n})}
- 1  ]  + ...\}  \ .  }
Thus the Born-Infeld action  appears  also
in the heterotic string theory but here it is
  a certain  combination
 of all-order   string loop
corrections ($F^{2n}$-term originates  from the
$e^{2(n-2)\p}$-term in the heterotic string loop expansion).

 This observation may be  useful, e.g.,   in
 looking for  heterotic string
solutions with large and approximately  constant
electromagnetic field\foot{One should probably also assume that
$e^\p$ is  large enough in order  to be able to drop other $F^n$-terms
present  in the heterotic action (like the tree-level
$\tr F^2 \tr F^2$-term in \attos\  the contribution of
which is not included in the Born-Infeld action \acov).}
and also in trying to understand  the  action of the
four-dimensional
$S$-duality  when
terms of higher orders in $\a'$ are included
(see in this connection \gibbb).

\bigskip

While this paper was in preparation,
 further evidence
in favour of  the heterotic -- type I duality was provided in \pow.

\bigskip

I would like to thank  E. Bergshoeff, A. Dabholkar, M. de Roo,
G. Gibbons and  E. Witten
for stimulating  discussions and correspondence.
I acknowledge also the support of PPARC,
EC grant SC1$^*$-CT92-0789
and NATO grant CRG 940870.

\vfill\eject
  \listrefs
\vfill\eject
\end